# NUCLEAR PHYSICS AT THE TESLA*HERA COMPLEX [1]

## S. Sultansoy

*Deutsches Elektronen-Synchrotron, Notke Str. 85, 22607 Hamburg, Germany*
*Gazi Üniversitesi, Fen-Edeb. Fak., Fizik Böl., 06500 Teknikokullar, Ankara, Türkiye*
*Elmler Akademiyası Fizika Institutu, H. Cavid prospekti 33, Bakı, Azerbaycan*

## Abstract

Construction of the TESLA linear electron-positron collider tangentially to the HERA proton ring will provide a number of new facilities for particle and nuclear physics research. In this paper main parameters and physics goals of eA, γA and FEL γA colliders, as well as fixed target experiments are discussed.

**HERA based eA collider.** Collisions of $E_e$=30 GeV electrons with different nuclei accelerated in HERA proton ring ($E_A$=Z·0.92 TeV) will give opportunity to investigate parton distributions in nuclear medium. Especially, exploration of the region of very small $x$ at $Q^2$>1 GeV$^2$ is very important.

**TESLA*HERA based eA collider.** In this case $E_e$=250 GeV electrons from TESLA will be used, which results in essential extension of the $x$-range.

**TESLA*HERA based γA collider.** Using the Compton backscattering of the laser beam off the electron beam from TESLA one can form the high-energy γ-beam ($E_\gamma$≈200 GeV). The γA collider will give a unique opportunity to investigate a number of important phenomena (including the region of extremely small $x_g$) in a best manner.

**TESLA*HERA based FEL γA collider.** Ultrarelativistic ions will see the TESLA FEL (Free Electron Laser) beam as a beam of MeV energy photons. This will give a unique opportunity to investigate "old" nuclear phenomena in rather unusual conditions.

**Fixed target experiments.** The scattering of the polarised electron and photon beams from TESLA on polarised nuclear targets will give opportunity to investigate the spin contents of nucleons.

**ELFE@DESY: An Electron Laboratory For Europe at DESY.** Using TESLA and HERA a high luminosity quasi-continuous electron beam ($E_e$=15÷25 GeV) suitable for nuclear physics experiments can be produced. In addition, using a 2660 A$^o$ laser a 25 GeV electron beam can be converted into a photon beam with energy up to 16 GeV.

---

# 1. Introduction

The Deutsches Elektronen-Synchrotron [1] (German Electron Synchrotron, **DESY**), which was founded in 1959, is one of the largest international research centers in the world. More than 3400 scientists from 380 universities and institutes from the 35 countries (from Armenia to USA) are participating in experiments performing at the unique electron-hadron collider **HERA** (fundamental researches) and synchrotron radiation source **HASYLAB** (applied researches). All countries, which will be joined European Community at the first stage (Hungary, Poland etc) are represented at different scientific collaborations. DESY's purpose is to conduct basic research in the natural sciences, with activity focused on:
- The investigation of the fundamental properties of matter in particle physics
- The use of synchrotron radiation in surface physics, materials science, chemistry, molecular biology, geophysics, and medicine
- The development, construction and operation of corresponding accelerator facilities.

The fundamental properties of matter are investigated with four large detectors in the HERA ring, namely, **H1**, **ZEUS**, **HERMES** and **HERA-B**. First two deal with lepton-proton collisions at very high center-of-mass energies and investigate physics processes at extremely small distances. The last two: HERMES (collisions of 27 GeV energy polarised electron beam with polarised nucleus target) and HERA-B (collisions of 920 GeV energy proton beam with nucleus target) perform researches in the field of nuclear physics, also.

Synchrotron radiation is produced at **DORIS III** (4.5 GeV energy $e^+$ beam) and **PETRA II** (12 GeV energy $e^-/e^+$ beams) storage rings. Some examples of research at the HASYLAB are:
- X-ray fluorescence analysis for trace element analysis of materials
- Investigations of catalytic processes with high time resolution
- X-ray absorption spectroscopy for determination of short-range atomic order
- X-ray small-angle scattering investigations of segregations in metals
- Powder diffractometry for analysis of polycrystalline materials
- Structural investigations of bio-molecules with atomic resolution
- X-ray imaging of the coronary arteries without the use of a catheter in the arterial system.

**The TESLA Project**. As a part of an international collaboration, DESY is planning the TESLA research facility [2] − a 33-kilometer-long, super-conducting electron-positron linear collider (center-of-mass energy in excess of 500 GeV) with integrated X-ray lasers (free electron lasers without highly reflecting mirrors for the generation of X-ray radiation with wavelengths down to 0.1 nanometers).

The TESLA Test Facility (**TTF**) is being used to develop and test the super-conducting acceleration structures and to test the new SASE principle of the X-ray laser. Afterwards, it will be extended to form the 300-meter-long **TTF-FEL**. This powerful tool for investigations in wide spectrum of scientific and technological fields, which will be ready in 2003, has following unique features:

- A peak power **ten thousand million times higher** than that of the most modern X-ray sources
- Light of every wavelength in the range between 20 and 6 nanometers
- Up to 80000 extremely short light flashes per second.

Below I consider different options of the **TESLA\*HERA** complex, which have particular importance from the viewpoint of nuclear science. Let me remind you, that nuclear properties are already investigated at DESY with HERA-B and HERMES detectors (see, for example, [3] and references therein).

## 2. HERA based eA collider

An acceleration of nucleus beams in the HERA proton ring will give an opportunity to investigate nuclear physics in collisions of 27 GeV energy electrons with $Z \times 920$ GeV energy nuclei [4]. An estimations show that achievable luminosity for this option is given by $L_{eA} \approx (1/A) \cdot 10^{30 \div 31} \text{cm}^{-2}\text{s}^{-1}$ for light and medium mass nuclei. An impressive list of important nuclear phenomena, which can be investigated at this option, is presented in a number of papers in [4].

## 3. TESLA\*HERA based eA collider

Collisions of electron beam from TESLA with nucleus beam from HERA will give opportunity to extend essentially the kinematics of lepton-nucleus interactions [5-8]. The main limitation for this option (as in previous case) comes from fast emittance growth due to intra beam scattering, which is approximately proportional $(Z^2/A)^2(\gamma_A)^{-3}$. In this case, the using of flat nucleus beams seems to be more advantageous because of few times increasing of luminosity lifetime. Nevertheless, sufficiently high luminosity can be achieved at least for light nuclei. For example, $L_{eC}=1.1 \cdot 10^{29} \text{cm}^{-2}\text{s}^{-1}$ for collisions of 300 GeV energy electron beams and Carbon beam with $n_C = 8 \cdot 10^9$ and $\varepsilon_C^N = 1.25\pi \cdot \text{mrad} \cdot \text{mm}$. This value corresponds to $L_{int} \cdot A \approx 10 pb^{-1}$ per working year ($10^7$ s) needed from the physics point of view [9]. Similar to the *ep* option, the lower limit on $\beta_A^*$, which is given by nucleus bunch length, can be overcome by applying a "dynamic" focusing scheme [10] and an upgrade of the luminosity by a factor 3-4 may be possible. For recent status see [11].

## 4. TESLA\*HERA based γA collider

In my opinion this is the most promising option of the TESLA⊗HERA complex, because it will give unique opportunity to investigate small $x_g$ region in nuclear medium. Indeed, due to the advantage of the real $\gamma$ spectrum heavy quarks will be produced via $\gamma g$ fusion at characteristic

$$x_g \approx \frac{4m^2_c(b)}{0.83 \times (Z/A) \times s_{ep}},$$

which is approximately $(2 \div 3) \cdot 10^{-5}$ for charmed hadrons.

As in the previous option, sufficiently high luminosity can be achieved at least for light nuclei. Then, the scheme with deflection of electron beam after conversion is preferable because it will give opportunity to avoid limitations from $\Delta Q_A$, especially for heavy nuclei. The dependence of luminosity on the distance between conversion region and interaction point for TESLA⊗HERA based $\gamma C$ collider is similar to that of the $\gamma p$ option [12]: $L_{\gamma C}=1.3 \cdot 10^{29}$ cm$^{-2}$s$^{-1}$ at $z=0$ and $L_{\gamma C}=10^{29}$ cm$^{-2}$s$^{-1}$ at $z=5$ $m$ with 300 GeV energy electron beam. Let me remind that an upgrade of the luminosity by a factor 3-4 may be possible by applying a "dynamic" focusing scheme. Further increasing of luminosity will require the cooling of nucleus beam in the main ring. Finally, very forward detector in $\gamma$-beam direction will be very useful for investigation of small $x_g$ region due to registration of charmed and beauty hadrons. Preliminary list of physics goals contains [6]:

- Total cross-sections to clarify real mechanism of very high energy γ-nucleus interactions
- According to the VMD, this machine will be also ρ-nucleus collider
- Formation of the quark-gluon plasma at very high temperatures but relatively low nuclear density
- Gluon distributions at extremely small $x_g$ in nuclear medium
- Investigation of both heavy quark and nuclear medium properties via heavy quarkonia
- Existence of multi-quark clusters in nuclesr medium etc.

## 5. TESLA*HERA based FEL γA collider

Colliding of TESLA FEL beam with nucleus bunches from HERA may give a unique possibility to investigate "old" nuclear phenomena in rather unusual conditions. The main idea is very simple [13]: ultra-relativistic ions will see laser photons with energy $\omega_0$ as a beam of photons with energy $2\gamma_A\omega_0$, where $\gamma_A$ is the Lorentz factor of the ion beam. For HERA $\gamma_A=(Z/A)\gamma_p=980(Z/A)$, therefore, the region $0.1 \div 10$ MeV, which is matter of interest for nuclear spectroscopy, corresponds to $0.1 \div 10$ keV lasers, which coincide with the energy region of TESLA FEL. The excited nucleus will turn to the ground state at a distance $l=\gamma_A \cdot \tau_A \cdot c$ from the collision point, where $\tau_A$ is the lifetime of the excited state in nucleus rest frame and $c$ is the speed of light. The huge number of expected events (~$10^{10}$ per day for 4438 keV excitation of $^{12}$C) and small energy spread of colliding beams ($\leq 10^{-3}$ for both nucleus and FEL beams) will give opportunity to scan an interesting region with ~1 keV accuracy.

## 6. Fixed target options

**TESLA-N**. This option assumes the use of one arm of the TESLA collider for a polarized fixed target experiment in parallel to the collider experiments. The main advantage will be polarization of both the 250 GeV energy electron beam (approx. 80% longitudinal polarization) and nuclear target, as well as very high luminosity (approx. 600 fb$^{-1}$). For more details, see [14].

**REGAS at TESLA**. This option will give unique opportunuty to measure polarized gluon distribution with unprecendent accuracy. The scheme of the proposed experiment looks as follows [15]. A circularly polarized laser beam with a photon energy $\omega_0 = 3.3$ eV (Cu15 laser) is scattered off the 250 GeV energy electrons provided by TESLA. As the result we obtain high energy (~200 GeV), highly polarized (~100% longitudinal polarization) and practically monochromatic beam of photons, which is scattered on polarized nuclear target (deuterated butanol). An investigation of heavy (charm and beaty) quarks pair production will give an opportunity to measure polarization of gluons practically without background.

## 7. ELFE@DESY

**An Electron Laboratory For Europe (ELFE) at DESY.** The ELFE proposal is supported by NuPECC (the Nuclear Physics European Collaboration Committee). ELFE will be the first high energy electron beam beyond 10 GeV with both high intensity and high duty factor. Construction of the TESLA at the DESY site will give opportunity to realize ELFE project at DESY site. Using part of TESLA as an injector (corresponding to 15÷25 energy electron beam), the achievable value of the current stored in the HERA ring is 150 mA [2, 16]. The general physics motivation presented at NuPECC Vienna meeting (April 1994) and included into the NuPECC recommendations is following:
- The investigation of strongly interacting systems with the elementary probe − the quark, produced in electron-quark scattering − is essential
- Studies of hadron structure by exclusive experiments are indispensible for a better understanding of QCD in the confinement region
- New windows for the investigation of hadronic matter are opened by the use of probes with strangeness and charm.

In addition, the laser backscattering technique offers the possibility of producing photon beams. For example, using a 2660 $A^o$ laser in combination with a 25 GeV electron beam available at DESY, a maximum laser backscattering photon energy of 16 GeV can be obtained.

## 8. Conclusion

Taking into acount both the existing and planned research instrumentation facilities, **Turkic states must join to existing and future International Collaborations at the DESY** as soon as possible.

**Acknowledgements.** I would like to express my gratitude to R. Brinkmann, A. Celikel, A.K. Ciftci, M. Kantar, F. Willeke, O. Yavas and M. Yilmaz for fruitful collaboration. Special thanks to M. Leenen and D. Trines for useful discussions and valuable remarks. I am grateful to Professor C. Yalcin for his kind invitation to participate the I. Eurasia Conference on Nuclear Science and Its Applications.
This work is supported by Turkish State Planning Organisation under the Grant No DPT-97K-120420 and DESY.